# Identification of Novel Diagnostic Neuroimaging Biomarkers for Autism Spectrum Disorder Through Convolutional Neural Network-Based Analysis of Functional, Structural, and Diffusion Tensor Imaging Data Towards Enhanced Autism Diagnosis

Annie Adhikary[1]

[1] Nikola Tesla STEM High School, 4301 228th Ave NE, Redmond WA 98053, USA

**Abstract.** Autism spectrum disorder is one of the leading neurodevelopmental disorders in our world, present in over 1% of the population and rapidly increasing in prevalence, yet the condition lacks a robust, objective, and efficient diagnostic. Clinical diagnostic criteria rely on subjective behavioral assessments, which are prone to misdiagnosis as they face limitations in terms of their heterogeneity, specificity, and biases. This study proposes a novel convolutional neural network-based classification tool that aims to identify the potential of different neuroimaging features as autism biomarkers. The model is constructed using a set of sequential layers specifically designed to extract relevant features from brain imaging data. Trained and tested on over 300,000 distinct features across three imaging types, the model shows promise in classifying individuals with autism from typical controls, outperforming metrics of current gold standard diagnostics by achieving an accuracy of 95.4% on a dataset of 1,111 samples with 521 autistic subjects (260 male and 261 female) and 590 controls (297 male and 293 female). 32 optimal features from the training data were identified and classified as candidate biomarkers using an independent samples t-test, in which functional features such as connectivity and the time series of signal intensity from each voxel exhibited the highest mean value differences between individuals with autism and typical control subjects. The p-values of these biomarkers were < 0.001, proving the statistical significance of the results and indicating that this research could pave the way towards the usage of neuroimaging in conjunction with behavioral criteria in clinics. Furthermore, the salient features discovered in the brain structure of individuals with autism could lead to a more profound understanding of the underlying neurobiological mechanisms of the disorder, which remains one of the most substantial enigmas in the field even today.

**Keywords:** Autism spectrum disorder, neuroimaging biomarkers, brain regions, diagnostics, magnetic resonance imaging, convolutional neural network, neurodevelopment.

# 1 Introduction

## 1.1 Background and Current Diagnostic Methods

This study aims to identify robust neuroimaging biomarkers for autism spectrum disorder (ASD) with the goal of developing a more precise and effective diagnostic tool that proves effective for both male and female individuals on the autism spectrum. ASD is a complex neurodevelopmental disorder currently lacking objective diagnostic tools, which can lead to inconsistencies in diagnosis and treatment. The existing gold standard, the Autism Diagnostic Observation Schedule, Second Edition (ADOS-2), suffers from a variety of limitations such as subjective evaluations by clinicians, an inability to accurately distinguish between varying symptom severities, and inadequate early detection capabilities as it primarily assesses existing conditions in children. Objective diagnosis of autism based on unbiased medical data, like neuroimaging, holds the potential to provide a more accurate and dependable assessment of an individual's condition compared to subjective diagnostic methods that are prone to bias and variability among clinicians. Non-invasive whole-brain scans may offer valuable insights for diagnosing neuropsychiatric disorders like autism, due to the neuroscientific nature of the disorder.

The assessment of ASD is filled with challenges due to the limitations of publicly available imaging datasets and the subjectivity of dominant diagnostic methods such as the ADOS-2 or the Diagnostic and Statistical Manual of Mental Disorders, 5th Edition (DSM-5). To address these challenges and enhance the accuracy and reliability of ASD diagnosis, recent research has turned to machine learning and deep learning techniques for identifying unique neural patterns in brain imaging data from individuals with autism. With the use of advanced computational methods to investigate potential patterns in the neural underpinnings of autism, this study aims to contribute to the development of a more reliable and specific diagnostic tool for individuals on the autism spectrum.

**Table 1.** Trends in autism diagnosis rates over the years as reported by the Centers for Disease Control and Prevention.

| Surveillance Year | Birth Year | Number of Sites Reporting | Combined Prevalence per 1,000 Children (Range Across Sites) | This is about 1 in X children |
| --- | --- | --- | --- | --- |
| 2020 | 2012 | 11 | 27.6 (23.1-44.9) | 1 in 36 |
| 2018 | 2010 | 11 | 23.0 (16.5-38.9) | 1 in 44 |
| 2016 | 2008 | 11 | 18.5 (18.0-19.1) | 1 in 54 |
| 2014 | 2006 | 11 | 16.8 (13.1-29.3) | 1 in 59 |
| 2012 | 2004 | 11 | 14.5 (8.2-24.6 | 1 in 69 |
| 2010 | 2002 | 11 | 14.7 (5.7-21.9) | 1 in 68 |
| 2008 | 2000 | 14 | 11.3 (4.8-21.2) | 1 in 88 |
| 2006 | 1998 | 11 | 9.0 (4.2-12.1) | 1 in 110 |
| 2004 | 1996 | 8 | 8.0 (4.6-9.8) | 1 in 125 |
| 2002 | 1994 | 14 | 6.6 (3.3-10.6) | 1 in 150 |
| 2000 | 1992 | 6 | 6.7 (4.5-9.9) | 1 in 150 |

The issue this study seeks to address is the challenge of accurately diagnosing ASD and overcoming the limitations of current methods, particularly those that depend on behavioral observations and psychological questionnaires, which are prone to resulting in false negatives and require significant time to administer. Given the rising prevalence of ASD (see Table 1), which is now estimated to affect more than 2% of children worldwide (Christensen et al., 2018), the need for efficient and reliable diagnostic methods has become increasingly urgent.

ASD is a complex and heterogeneous disorder characterized by a variety of symptoms, including social impairments, repetitive behaviors, and difficulties with speech and communication. Due to its heterogeneity and wide spectrum, ASD often presents diagnostic challenges that may result in delayed or inaccurate identification (Hus & Segal, 2021). Traditional evaluation methods rely heavily on behavioral observations and psychological assessments, which are not only time-consuming but also subject to potential inaccuracies (Hyman et al., 2020). Given the increasing ASD prevalence and the limitations of existing diagnostic approaches, there is a pressing need to explore alternative methods that can more efficiently and effectively identify individuals on the autism spectrum. By focusing on the development of innovative diagnostic tools that harness advanced computational techniques and integrate multiple sources of data, this research aims to contribute to the optimization of ASD diagnosis and, ultimately, the enhancement of support and treatment options for those affected by the disorder.

## 1.2 Neuroimaging Biomarkers: What is Known

A diagnostic neuroimaging biomarker is a specific type of neuroimaging biomarker that is used to identify the presence of a particular disease or condition in an individual. It is a measurable feature detected through neuroimaging techniques, such as magnetic resonance imaging (MRI), that can differentiate between healthy individuals and those with the disease or condition of interest. Neuroimaging biomarkers have emerged as a promising area of research for understanding the neural correlates of neurodevelopmental disorders such as ASD. These biomarkers involve a variety of imaging techniques that provide valuable insights into the structural and functional organization of the brain and its role in the presence of ASD.

Functional MRI (fMRI) has been widely used to investigate functional connectivity between brain regions in individuals with ASD. By measuring changes in blood flow as an indicator of brain activity, fMRI could identify abnormal patterns of connectivity that may significantly influence the core symptoms of the disorder. Structural MRI (sMRI) examines brain morphology, such as gray and white matter volumes, cortical thickness, and surface area, and has been successfully used to detect structural differences in the brains of individuals with neurological disorders and diseases (Coluzzi et al., 2023). Comparing these structural features across groups could enable the identification of potential biomarkers associated with ASD. Diffusion MRI (dMRI) is another essential imaging technique that assesses the integrity of white matter by measuring water diffusion along axonal tracts. Previous studies investigating dMRI have revealed abnormalities in white matter microstructure and connectivity in individuals with ASD, offering further insights into the neural substrates of the disorder.

As research progresses, the combination of multiple imaging modalities, such as fMRI, sMRI, and dMRI, the most widely available and prevalent imaging methods in the field of neuroimaging, has immense potential to provide a comprehensive view of the structural and functional organization of the brain in individuals with ASD. This integration of techniques and tools can, therefore, lead to the development of an undeniably accurate and reliable diagnostic tool, which in turn produces targeted interventions for those affected by the disorder at hand. Recent advancements in computational methods, specifically neural networks, may also enable the analysis of large amounts of complex neuroimaging data, potentially leading to the discovery of novel biomarkers that were not identifiable before.

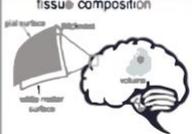

**Fig. 1.** Comparative analysis of the techniques, measures, and functionality of the brain imaging protocols employed in this study.

**Why Neuroimaging Biomarkers are Relevant.** Neuroimaging biomarkers are highly relevant in the context of neurodevelopmental disorders, such as ASD, because they provide a unique window into the underlying neural mechanisms that contribute to the wide range of symptoms and behaviors associated with these conditions. Traditional diagnostic methods, which rely heavily on subjective behavioral observations and psychological assessments, can be time-consuming, less reliable, and subject to inaccuracies. Neuroimaging biomarkers, on the other hand, offer a more objective, quantifiable, and data-driven approach to understanding the complex biological underpinnings of neurodevelopmental disorders.

The identification of reliable neuroimaging biomarkers has the potential to revolutionize the diagnosis and treatment of ASD. By providing a more accurate and earlier diagnosis, these biomarkers can enable timely and targeted interventions, leading to improved outcomes for affected individuals and their families. In addition, the discovery of specific neural correlates associated with these disorders can guide the development of novel, personalized therapeutic strategies that consider the unique neurobiological profile of an individual.

Neuroimaging biomarkers have the potential to significantly advance our understanding of the etiology and pathophysiology of neurodevelopmental disorders such as ASD. Unraveling the neural networks of the brain and their alterations in certain disorders may facilitate the identification of novel targets for intervention and prevention strategies. Furthermore, the integration of neuroimaging data with other sources of information, such as clinical diagnostic evaluations, can lead to a more comprehensive understanding of the complex interconnection between multiple determinants of neurodevelopmental disorders. The relevance of neuroimaging biomarkers in the study of neurodevelopmental disorders lies in their ability to provide objective, data-driven insights into the neural substrates of these conditions. This information holds tremendous potential for improving diagnostic accuracy, guiding personalized treatment approaches, and ultimately enhancing the quality of life for individuals affected by these disorders and their families.

**Challenges with Neuroimaging Biomarkers.** Despite the immense potential of neuroimaging biomarkers, there are several challenges that should be acknowledged and addressed. One of the primary issues is the heterogeneity of ASD itself, which manifests in diverse symptoms, cognitive profiles, and underlying genetic factors. This complexity makes it difficult to establish universally applicable biomarkers, as they may not capture the full range of variability present in the disorder.

Another challenge arises from the variability in neuroimaging data acquisition and processing. Different imaging sites employ varying imaging protocols, hardware, and analytical techniques, leading to inconsistencies in the findings and making it difficult to establish reliable and reproducible biomarkers. Standardization of these protocols and the development of harmonization techniques can help mitigate this issue and enhance the comparability of results across different studies.

The size and representativeness of study samples also pose significant challenges. Many neuroimaging studies have relatively small sample sizes, which can limit the generalizability of their findings. ASD research often underrepresents certain demographic groups, such as females and individuals from diverse cultural and socioeconomic backgrounds. Ensuring more inclusive and representative study samples is crucial for the identification of robust neuroimaging biomarkers that are applicable to the broader ASD population.

The integration of multimodal data and the application of advanced computational methods, such as machine learning and deep learning, present both opportunities and challenges. While these approaches hold great promise for the discovery of novel biomarkers and the refinement of our understanding of the neural underpinnings of ASD, they also require the development of appropriate analytical frameworks and the careful interpretation of results to avoid overfitting and falsified correlations. Addressing these challenges in the context of neuroimaging biomarkers for ASD is essential for improving diagnostic accuracy, guiding personalized treatment approaches, and advancing our understanding of the complex neural mechanisms underlying this heterogeneous disorder.

**Past Research.** The application of machine learning and deep learning techniques, particularly convolutional neural networks (CNNs), in neuroimaging research has shown potential in identifying biomarkers and predicting autism based on MRI data (Stember et al., 2022). However, despite these promising developments, there remain several challenges and limitations that warrant further investigation and refinement.

Many of the previously developed models using machine learning and deep learning techniques have limitations in their performance metrics, including accuracy, precision, recall, and F1-score (Halibas et al., 2018). These limitations may be partly attributed to the fact that most previous models have only considered shallow features from MRI data, rather than extracting more complex, higher-level features that may be more indicative of autism (Anagnostou & Taylor, 2011). Furthermore, the heterogeneity of ASD poses significant challenges for the identification of reliable and generalizable neuroimaging biomarkers. Many studies have focused on specific subgroups or aspects of autism, making it difficult to draw overarching conclusions about the disorder. Additionally, research samples are often small and lack representation from diverse demographic groups, limiting the generalizability of findings (Parellada et al., 2022).

Another critical issue is the need for standardization of data acquisition and processing methods across studies. Variability in imaging protocols may additionally lead to inconsistencies in results, hindering the establishment of reliable and reproducible biomarkers (A et al., 2020). While advanced computational methods such as deep learning have demonstrated the ability to extract complex patterns from neuroimaging data, they also present challenges in terms of interpretability and potential underfitting or overfitting. Ensuring the validity and robustness of these methods requires the development of appropriate analytical frameworks and cautious interpretation of results (Loth et al., 2016). Although machine learning and deep learning techniques have made significant contributions to the field of neuroimaging biomarkers for ASD, there remains ample room for improvement and further exploration. Novel research that addresses these challenges and limitations can enhance our understanding of ASD's neural substrates and contribute to the development of more reliable, specific, and inclusive diagnostic tools for individuals on the autism spectrum.

## 1.3   Study Goals

To address these limitations and improve the accuracy of autism diagnosis, this study aims to build a novel deep learning model involving the use of 3D MRI data and preprocessing techniques such as slice time correction and normalization to extract features from a lower level to a higher level and improve classification accuracy for autism. By training the model on this data, the model aims to achieve an accuracy of over 95% in classifying autism versus typical controls, a significant improvement over previous solutions that have struggled to achieve similar levels of accuracy.

To address the exclusion of women in previous ASD research and improve classification accuracy, this model will incorporate data from equal populations of males and females with ASD and apply statistical techniques to identify features specific to each sex. It is important to use both male and female data in studies because sex can have significant effects on a wide range of biological, psychological, and social phenomena. This is particularly important in fields such as neuroimaging, where there may be significant differences in brain structure and function between males and females (Ruigrok et al., 2014). By including data from both males and females, this study will help to identify and control for these differences, ensuring that the findings from this study are more representative of the general population and are not biased towards one sex such as existing solutions, including ADOS-2, that were developed using a largely male sample and consequently make it more difficult to develop useful interventions or provide accurate diagnoses for girls and women (Skogli et al., 2013).

To address the biases and limitations in the current diagnostic methods, it is crucial to develop a more objective, specific, and functional diagnostic tool for ASD. To achieve this, this study aims to ensure that the model is trained on a diverse and representative dataset that includes a balanced representation of both ASD and typical control (TC) subjects. The dataset aims to encompass a variety of individuals, with equal proportions of males and females and adequate representation across different cultural and linguistic backgrounds. By incorporating diversity within the dataset, the model aims to capture the variability in the general population, effectively avoiding overfitting to certain demographic subgroups.

In addition to creating objectivity, this study aims to enhance the specificity of the diagnostic tool created by leveraging novel, in-depth neuroimaging analysis techniques. This approach allows for the extraction of a comprehensive set of features that can accurately differentiate ASD from TC subjects. By examining many features derived from three types of neuroimaging data, the model aims to provide a more nuanced and specific understanding of the neurobiological underpinnings of ASD. The diagnostic tool also aims to analyze the extracted neuroimaging features effectively. If successfully analyzed, the model results will help identify candidate diagnostic biomarkers for ASD, which can then be employed to improve the accuracy, specificity, and objectivity of ASD diagnostics. Ultimately, these advancements will lead to better identification and support for individuals affected by ASD.

## 2 Methodology

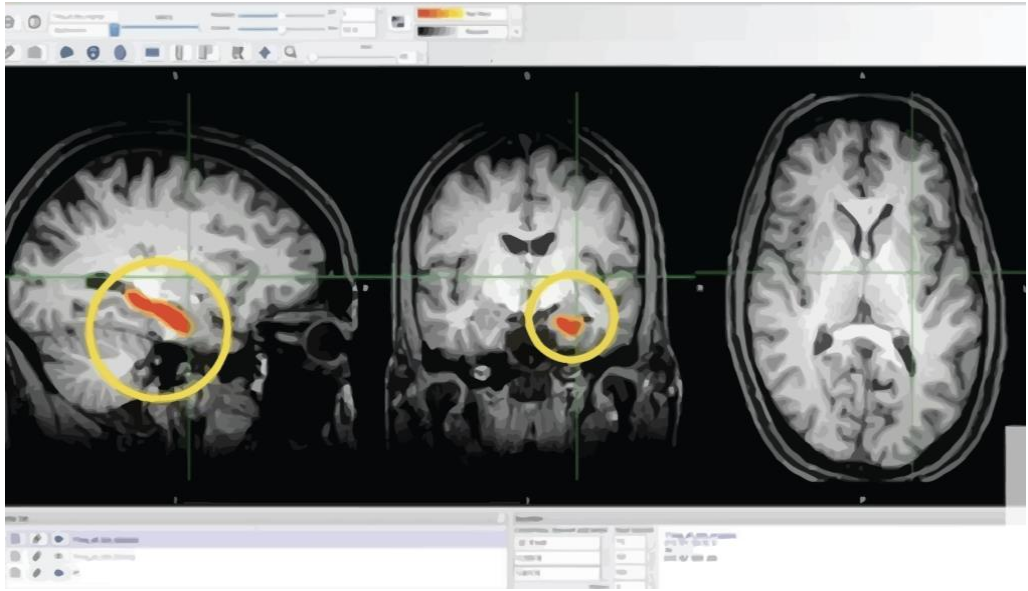

**Fig. 2.** Metrics extracted from imaging data sourced from the Autism Brain Imaging Data Exchange.

### 2.1 Aggregating and Understanding the Data

**Collecting Magnetic Resonance Images.** The raw MRI data utilized in this study was sourced from the Autism Brain Imaging Data Exchange I (ABIDE-I) dataset, obtained via the Connectome Computation System (CCS) pipeline. As a publicly accessible dataset, ABIDE-I is widely recognized and utilized by the neuroimaging community, containing a substantial sample of subjects diagnosed with ASD as well as typical controls (TC). The focus of this investigation is centered on data retrieved across seven sites, which were chosen from 19 available sites within the dataset. These sites consist of studies from the California Institute of Technology, Carnegie Mellon University, Kennedy Krieger Institute, Leuven Autism Research Consortium, Oregon Health & Science University, New York University, Munich, and others. To ensure consistency and comparability across the entire sample, the CCS ABIDE data underwent preprocessing using standardized pipelines that were applied uniformly to all subjects. The initial stage of data processing involved loading the three-dimensional (3D) MRI data and converting it into two-dimensional (2D) images.

Slice time corrections and normalizations were implemented to normalize the data, effectively adjusting the images to meet the necessary specifications for further analysis. The use of standardized preprocessing pipelines and parameters is a crucial component of the study's methodology, as it guarantees that any variations observed in the data are attributable to the inherent biological differences between subjects rather than disparities in the preprocessing steps. In addition, necessary steps were taken on each of the three distinct types of imaging used to prepare them for analysis. By adhering to these meticulous standards, this study ensures the validity of the data and the validity of the subsequent analyses, thus contributing valuable insights into the complex world of neuroimaging and our understanding of autism.

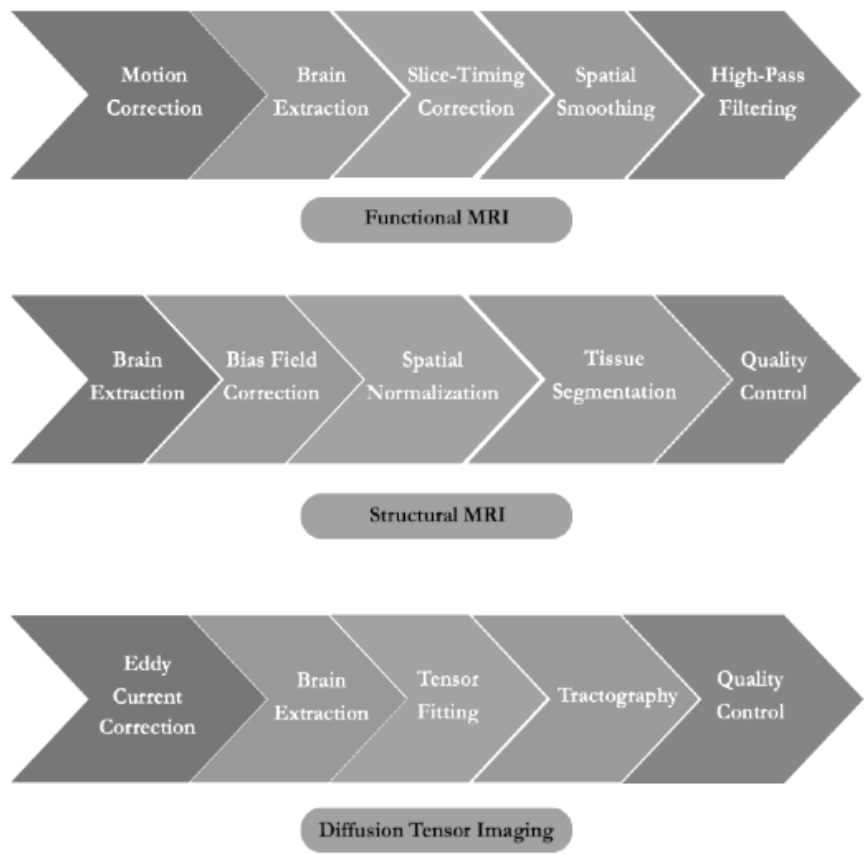

**Fig. 3.** Distinct preprocessing procedures applied to each type of imaging data using the FMRIB Software Library.

**Data Preprocessing.** A total of 1,111 samples was aggregated, drawing from an extensive array of 19 university study databases. For each subject, one fMRI scan, one sMRI scan, and one dMRI scan was collected. This dataset comprised a total of 3,333 neuroimaging files, one of each type of imaging per sample, creating a large and diverse dataset that significantly expands on the number of patients used in previous neuroimaging studies concerning ASD. A nearly equal amount of ASD and TC (521 and 590 respectively) samples were selected along with a nearly equal amount of male and female (557 and 554 respectively) samples, resulting in a representative selection of subjects.

To maintain consistency and optimize the comparability of results across various imaging modalities, a unique preprocessing pipeline has been employed for each type of neuroimaging data. This standardization process ensures that the data are normalized and standardized, allowing for more accurate and reliable analyses. The widely used FMRIB Software Library (FSL) was utilized in this procedure, providing a robust and efficient platform for visualizing and preprocessing the neuroimaging scans. Access to FSL was facilitated with a terminal interface, streamlining the overall processing workflow.

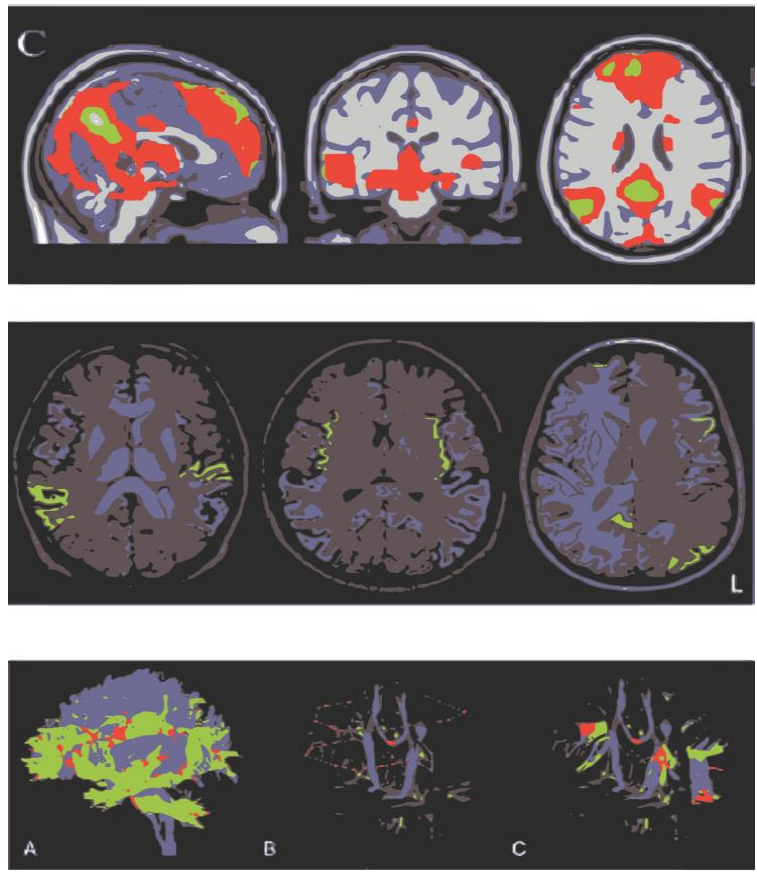

**Fig. 4.** Highlighted extracted features within standardized (functional, structural, diffusion) brain images.

**Feature Extraction.** The application of mathematical algorithms in the field of neuroimaging analysis has revolutionized the way we can interpret and understand complex brain activity patterns. Among these tools is the computational library SimpleITK, which uses advanced algorithms to rapidly extract hundreds of features from various aspects of neuroimages, such as activity in specific brain regions and the connectivity between these areas. This comprehensive extraction process enables a more in-depth investigation into the intricacies of brain function and the underlying mechanisms of various neurological disorders.

Following the extraction of an extensive array of features from the neuroimaging data, it is crucial to identify the most relevant and informative features for inclusion in the analytical model. This step is essential to minimize the risk of overfitting or underfitting the model, which could lead to inaccurate predictions and hinder the overall effectiveness of the analysis. To accomplish this, the Recursive Feature Elimination (RFE) technique was employed, a widely used and efficient algorithm that systematically determines the most informative features for a given classification problem.

The RFE algorithm accomplishes this task by iteratively evaluating the importance of each feature and removing the least significant ones, ultimately refining the feature set to include only those that contribute the most to the model's predictive power. Based on the feature rankings generated by the RFE algorithm, a decision was made to utilize the top 106 most important features from each imaging type for the development of the analytical model as they were among the top 10% of features as ranked by the algorithm. This selection process ensures that the model is both robust and reliable, maximizing its potential to provide valuable insights into the complex world of neuroimaging data and contribute to our understanding of brain function and dysfunction.

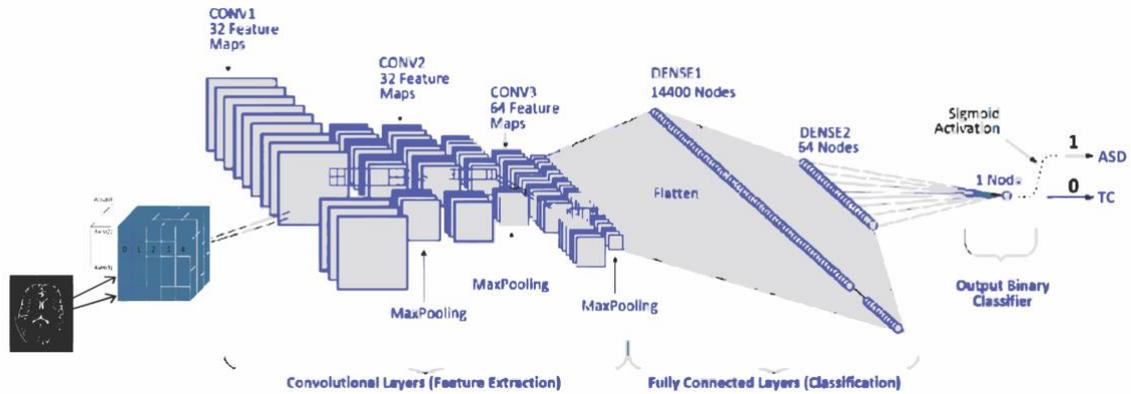

**Fig. 5.** Convolutional neural network processing input feature values to generate a predictive output.

## 2.2 Deep Learning

**Development, Training, and Testing of Models.** The deep learning methodology employed in the model building process is a crucial aspect of any research study, as it serves as the foundation upon which the validity and reliability of the results are built. In the present analysis, a total of 777 samples were utilized for the construction of the model, comprising approximately 70% of the entire dataset. This sizable subset of the data was carefully selected to ensure that the model is both representative and generalizable, thus increasing the likelihood of obtaining accurate and meaningful results.

The independent variables in the model consisted of neuroimaging features extracted from the various types of scans. These features provided a comprehensive overview of the different aspects of brain activity and connectivity, allowing for a more in-depth investigation into the relationships between these variables and the dependent variable, the diagnostic group to which each subject belongs.

To enhance the robustness and reliability of the model, the analysis was iterated three times, each time utilizing different input values derived from the fMRI, sMRI, and dMRI scans. This iterative approach allowed for a more thorough evaluation of the model's performance across various combinations of input values, ensuring that the final model is well-equipped to handle the complexities and nuances of the neuroimaging data. By incorporating a large sample size, a diverse array of neuroimaging features, and an iterative model-building process, the methodology employed in this study was designed to maximize the potential for obtaining meaningful and actionable insights. These results not only contribute to our understanding of the underlying mechanisms governing brain function and dysfunction but also hold promise for the development of more accurate diagnostic tools and targeted interventions for individuals affected by neurological disorders.

In this study, the performance of the proposed models was thoroughly assessed using four benchmark metrics, providing a comprehensive evaluation of the models' effectiveness in classifying subjects based on their neuroimaging data. To ensure the validity and reliability of the results, a 10-fold cross-validation procedure was conducted, which involved partitioning the data into ten equal subsets and iteratively training and testing the models on these subsets.

$$E = \frac{1}{10}\Sigma_{i=1}^{10} E_i \qquad (1)$$

The objective of the analysis was to identify the best performing imaging type for classification, with a focus on the average accuracy, F1-score, precision, and recall. These metrics were carefully chosen to provide a well-rounded assessment of the models' performance, considering factors such as the proportion of correct predictions (accuracy), the balance between precision and recall (F1-score), the ability to correctly predict the ASD class (true positives) and typical controls (true negatives), and the rates of incorrect predictions for ASD (false positives) and typical controls (false negatives). An in-depth search was conducted to learn the optimal hyperparameters for each model, leading to numeric results (see Table 2).

**Table 2.** Performance metrics of the model tested on different modalities of brain imaging data.

| Performance Measures | fMRI   | sMRI   | dMRI   |
| --- | --- | --- | --- |
| Accuracy  | 95.40% | 74.70% | 69.30% |
| Precision | 94.50% | 73.60% | 68.90% |
| Recall    | 92.50% | 73.30% | 66.50% |
| F1-Score  | 95%    | 74.50% | 69.10% |
| AUC Score | 94.70% | 74.33% | 68.93% |

**Results.** The evaluation of model performance metrics suggests that fMRI may be the most informative type of neuroimaging data for autism classification. This finding underscores the potential of fMRI as a valuable diagnostic tool in the field of autism research and clinical practice. Consequently, further in-depth analysis was conducted on the highest performing model, focusing on the fMRI data.

Several methods were employed to better understand and evaluate the performance of the fMRI model, including a confusion matrix. This matrix provides a summary of the model's performance by displaying the counts of true positives (TP), false positives (FP), true negatives (TN), and false negatives (FN). By presenting these values in a comprehensive format, the confusion matrix offers valuable insights into the model's ability to correctly classify subjects and identify potential areas for improvement. In addition, a receiver operating characteristic (ROC) curve was used to model the metrics of the model. The ROC curve illustrates the trade-off between the TP rate and the FP rate as the threshold for classification is varied. This graphical representation allows us to assess the overall performance of the model across different classification thresholds and determine the optimal balance between sensitivity and specificity. A train versus validation accuracy model was also employed. This analysis involves examining the trend of the model's accuracy during the training and testing phases. The training accuracy measures how well the model fits the training data, while the validation accuracy evaluates the model's ability to generalize to new, unseen data. By comparing these two measures, future research can better understand the model's performance and identify potential issues such as overfitting or underfitting.

Based on the evaluation of the five benchmark metrics, it is evident that the fMRI model outperforms most similar models in the field, but additionally has the potential to improve upon the metrics of clinical practices in theory if the results remain constant throughout larger independent datasets. 10-fold-cross-validation was additionally performed to validate, in theory, how the model would perform on an independent dataset, and the results remained largely constant with a variance in metrics less than 1%.

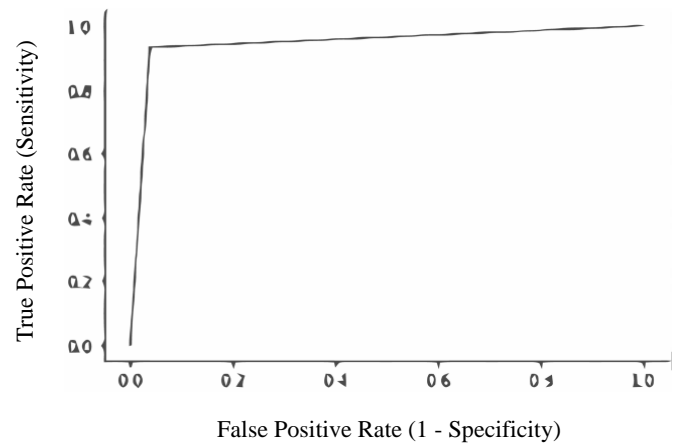

False Positive Rate (1 - Specificity)

**Fig. 6.** Receiver operating characteristic curve of fMRI model generated through TensorFlow software. High curve correlates with increased model accuracy in classification tasks.

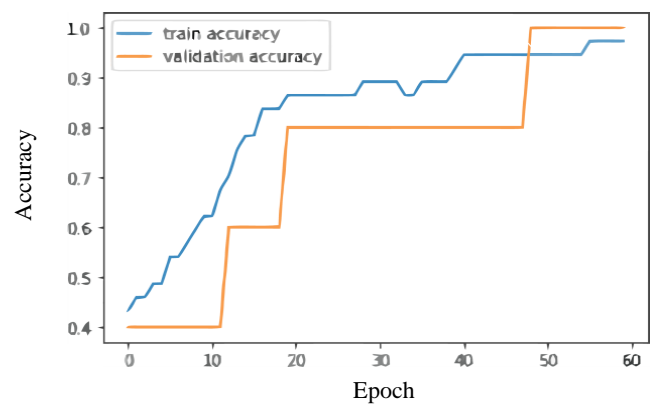

Epoch

**Fig. 7.** Train accuracy versus validation accuracy of fMRI model generated through TensorFlow software. Close alignment suggests the model's reliability and generalizability to unseen data.

## 3   Data Analysis and Results

### 3.1   Statistical Analyses for Biomarker Selection

In this study, the optimal features extracted from the functional imaging data were identified using independent sampling t-tests, a statistical method designed to compare the means of two groups. The top 32 features exhibiting the highest differences in mean values between individuals with autism and typical control subjects were chosen as candidate biomarkers as these potential biomarkers exhibited p-values less than 0.001 (Kennedy-Shaffer, 2017), indicating a high degree of statistical significance and supporting their relevance in distinguishing between the two groups. The candidate biomarkers include increased amygdala activation during emotion processing tasks, heightened default mode network connectivity during resting states, and altered motor network coherence during movement-related tasks. Furthermore, this study observed marked differences in the frontal lobe's functional connectivity, particularly within the dorsolateral prefrontal cortex, which is crucial for executive function, and aberrations in the visual cortex during stimulus processing. This study also identified potential biomarkers in the temporal lobes, where aberrant auditory processing was evident in the superior temporal gyrus, suggesting atypical integration of sound and language. Reduced connectivity in the fusiform gyrus, associated with facial recognition deficits, emerged as another neural signature of autism. Additionally, altered effective connectivity between the parietal lobes and prefrontal cortex was found, impacting attention and sensory integration. Variations in hemodynamic responses within the salience network, governing the detection of important stimuli, were also pronounced. Dynamic functional connectivity measures revealed fluctuations in neural synchrony, especially within the executive-control network, signifying irregular network stability. Increased local synchronization, as measured by Regional Homogeneity (ReHo), particularly in the insula and adjacent operculum, was also significant. Voxel-Mirrored Homotopic Connectivity (VMHC) findings suggest a reduced interhemispheric coordination, particularly between the temporal regions, which may underpin the social communication difficulties characteristic of autism. Network centrality measures within key regions implicated in social cognition, such as the anterior cingulate cortex and orbitofrontal cortex, displayed substantial divergence from control subjects. Degree centrality within the sensorimotor network also featured prominently, indicating potential motoric discrepancies. Further, the Granger Causality Analysis pointed towards atypical information flow from the occipital to frontal areas, which may influence visual information processing.

All identified biomarkers were from the computationally extracted set of 106 functional features evaluated for each individual in the study.

## 3.2  Male and Female Layer Separation

Notably, the analysis revealed distinct biomarkers for each sex group, underscoring the importance of considering sex differences when examining the neural underpinnings of autism. For biological females, pivotal biomarkers included increased amygdala activation during emotional tasks, heightened default mode network connectivity in resting states, altered coherence in the motor network during movement, aberrant auditory processing in the superior temporal gyrus, and atypical integration of visual stimuli in the visual cortex. For biological males, prominent biomarkers featured irregular activation in the dorsolateral prefrontal cortex impacting executive function, disrupted connectivity in the frontal lobe, particularly within the executive-control network, reduced interhemispheric coordination as evidenced by VMHC, especially in the temporal regions, and abnormalities in the salience network's hemodynamic response, alongside atypical Granger Causality information flow from the occipital to frontal areas. In addition to these sex-specific biomarkers, 22 other features were identified for each sex group, further emphasizing the unique neural characteristics associated with autism in males and females. By retrieving features and metrics separately for data labeled as male and female, this study was able to identify distinct biomarkers for each sex group. The highly differential results between groups of this approach not only highlights the importance of considering sex differences in autism research but also holds promise for the development of more targeted and effective diagnostic tools and interventions tailored to the unique needs of individuals affected by ASD.

## 3.3  Differences in Imaging Protocols

There was an observed slight variation in mean metrics between datasets, which highlights the potential influence of differences in imaging protocols between sites on the results. To combat this issue, future studies should investigate the effect of the methodology used for getting MRI scans on the results and accuracy of resulting biomarkers, leading to a potential standardized approach for medical imaging collection.

## 4  Conclusions and Discussion

The study presented several key findings and implications. Among its accomplishments included the extraction of 106 neuroimaging features from 1,111 unique patients and the successful assembly of a robust dataset suitable for deep learning applications. The initial study demonstrated that the proposed model achieved an impressive 95.4% accuracy, outperforming previous statistical analyses conducted on the DSM-5 and ADOS-2 diagnostics. Moreover, t-tests suggested that fMRI might be the most informative type of neuroimaging data for autism classification, offering valuable insights for future research and clinical practice. The study's findings have significant implications for various aspects of autism research and treatment. For instance, early detection and diagnosis can be greatly improved by using the identified candidate biomarkers. Detecting autism at an early stage is crucial for enhancing treatment outcomes and improving the quality of life for affected individuals and their families. Furthermore, the identification of neuroimaging biomarkers for autism can contribute to a better understanding of the underlying biology of the condition. This increased understanding can, in turn, lead to the development of novel treatments and therapies that target the root causes of autism. Additionally, neuroimaging biomarkers can be employed to identify patient subgroups for clinical trials, which can enhance the efficiency and effectiveness of these trials. This can result in faster development of new treatments and therapies for autism, ultimately benefiting those affected by the condition and their support networks.

I would like to thank my mentors: Dr. Jill E. Howard, Assistant Professor in Psychiatry and Behavioral Sciences at Duke University's Child & Family Mental Health & Community Psychiatry Division, and Mr. Silas Busch, PhD Candidate at the University of Chicago, for their generous critiques, comprehensive answers to my inquiries, and unwavering support of my research undertakings in recent months. I would also like to thank Ms. Kate Allender of Nikola Tesla STEM High School for equipping me with the necessary resources for rigorous scientific exploration and consistently nurturing my intellectual pursuits with unwavering encouragement.